\documentclass{emulateapj}

\usepackage{amssymb}
\usepackage{amsmath}
\usepackage{graphicx}
\usepackage{natbib}
\usepackage{txfonts}

\bibpunct{(}{)}{;}{a}{}{,} 

\def\roma{1}
\def\icra{2}
\def\exeter{3}
\def\ice{4}
\def\ieec{5}
\def\upc{6}

\shorttitle{A white dwarf merger as progenitor of the AXP 4U~0142+61?}
\shortauthors{Rueda et al.}

\begin{document}
\title{A  white dwarf  merger  as progenitor  of  the anomalous  X-ray
  pulsar 4U 0142+61?}

\author{J. A. Rueda\altaffilmark{\roma,\icra}, 
        K. Boshkayev\altaffilmark{\roma,\icra}, 
        L. Izzo\altaffilmark{\roma,\icra},
        R. Ruffini\altaffilmark{\roma,\icra},
        P. Lor\'en--Aguilar\altaffilmark{\exeter},
        B. K\"ulebi\altaffilmark{\ice,\ieec},
        G. Aznar--Sigu\'an\altaffilmark{\upc,\ieec} \&
        E. Garc\'\i a--Berro\altaffilmark{\upc,\ieec}
}

\altaffiltext{\roma}{Dipartimento di Fisica and ICRA, 
                     Sapienza Universit\`a di Roma, 
                     P.le Aldo Moro 5, 
                     I--00185 Rome, 
                     Italy}

\altaffiltext{\icra}{ICRANet, 
                     P.zza della Repubblica 10, 
                     I--65122 Pescara, 
                     Italy}

\altaffiltext{\exeter}{School of Physics, 
                       University of Exeter, 
                       Stocker Road, 
                       Exeter EX4 4QL,
                       United Kingdom}

\altaffiltext{\ice}{Institut de Ci\`encies de l'Espai (CSIC), 
                    Facultat de Ci\`encies, 
                    Campus UAB, 
                    Torre C5-parell, 
                    08193 Bellaterra, 
                    Spain}

\altaffiltext{\ieec}{Institute for Space Studies of Catalonia, 
                     c/Gran Capit\`a 2--4, 
                     Edif. Nexus 104, 
                     08034 Barcelona, 
                     Spain}

\altaffiltext{\upc}{Departament de F\'isica Aplicada, 
                    Universitat Polit\`ecnica de Catalunya, 
                    c/Esteve Terrades, 5, 
                    08860 Castelldefels, 
                    Spain}

\altaffiltext{}{jorge.rueda@icra.it, enrique.garcia-berro@upc.edu}

\begin{abstract}
It   has   been   recently   proposed   that   massive   fast-rotating
highly-magnetized  white  dwarfs   could  describe  the  observational
properties of  some of Soft  Gamma-Ray Repeaters (SGRs)  and Anomalous
X-Ray  Pulsars  (AXPs).   Moreover,  it   has  also  been  shown  that
high-field magnetic (HFMWDs) can be  the outcome of white dwarf binary
mergers.  The products of these mergers consist of a hot central white
dwarf surrounded  by a rapidly rotating  disk.  Here we show  that the
merger of a  double degenerate system can  explain the characteristics
of  the  peculiar  AXP  4U~0142+61. This  scenario  accounts  for  the
observed infrared excess.   We also show that  the observed properties
of 4U~0142+6 are consistent with  an approximately $1.2~M_{\sun}$ white 
dwarf,
remnant of  the coalescence of  an original  system made of  two white
dwarfs of masses  $0.6\, M_{\sun}$ and $1.0\,  M_{\sun}$.  Finally, we
infer a post-merging age $\tau_{\rm WD}\approx 64$~kyr, and a magnetic
field $B\approx 2\times  10^8$~G.  Evidence for such  a magnetic field
may  come  from  the  possible detection  of  the  electron  cyclotron
absorption feature observed between the  $B$ and $V$ bands at $\approx
10^{15}$~Hz in the spectrum of 4U~0142+61.
\end{abstract}


\maketitle

\section{Introduction}

SGRs are  sources of short  ($\approx 100$~ms), repeating bursts  of soft
$\gamma$-ray  and X-ray  radiation at  irregular intervals,  and share
with  AXPs  several  similarities,  like  rotation  periods  clustered
between 2 and 12~s, and high magnetic fields.  Their observed spindown
rates range  from $\dot{P}  \approx (10^{-15}$  to $10^{-10})$,  and have
typical    X-ray   luminosities    in    quiescent   state    $L_X\approx
10^{35}$~ergs$^{-1}$.   Currently, it  is widely  accepted that  these
objects are magnetars  \citep{duncan92,thompson95}, although there are
competing scenarios that  challenge this model ---  see, for instance,
the excellent  and recent  review of \cite{sandrorev},  and references
therein.    Recently,    \cite{2012PASJ...64...56M},   following   the
pioneering works of \cite{1988ApJ...333..777M} and \cite{paczynski90},
have suggested an alternative model that could explain some properties
of these sources. This  model involves highly-magnetized white dwarfs.
For this model to be viable the  masses of the white dwarfs need to be
rather  large ($M\ga  1.2\, M_{\sun}$),  their magnetic  fields should
range from  $B \approx  10^7$~G all  the way  to $10^{10}$~G,  and the
rotation  periods should  be  rather  small, of  the  order  of a  few
seconds.   The most  apparent drawback  of this  scenario, namely  the
rotational stability of fast rotating  white dwarfs, has been recently
analyzed.   Specifically, the  crucial  question  of whether  rotating
white dwarfs can  have rotation periods as short as  the ones observed
in AXPs has been recently addressed by \cite{2013ApJ...762..117B}, who
found that the minimum rotation  period of typical carbon-oxygen white
dwarfs is approximately $0.5$~s.  Thus, since AXPs have rotation periods larger
than this value they could be white dwarfs.

The  existence  of white  dwarfs  with  magnetic fields  ranging  from
$10^7$~G  up  to  $10^{9}$~G  is  solidly  confirmed  by  observations
\citep{2009A&A...506.1341K}.  Observations  show that most  HFMWDs are
massive, and moreover, that none  of them belongs to a non-interacting
binary system,  pointing towards  a binary origin.   However, although
long-suspected \citep{2000PASP..112..873W}, it  has only been recently
shown that  HFMWDs might be  the result  of white dwarf  mergers.  SPH
simulations of the coalescence process indicate that the result of the
merger is  a white  dwarf that  contains the  mass of  the undisrupted
primary, surrounded by a hot corona made  of about half of the mass of
the disrupted  secondary.  In  addition, a rapidly  rotating Keplerian
disk which contains the rest of  the material of the secondary is also
formed,  as  little  mass  is  ejected  from  the  system  during  the
coalescence process.   The rapidly-rotating  hot corona  is convective
and an efficient $\alpha\omega$ dynamo  can produce magnetic fields of
up to $B\approx 10^{10}$~G \citep{enrique2012}.

In view of these considerations it is natural to ask ourselves if such
binary mergers could  also explain the properties of  some AXPs.  Here
we explore such possibility for the  specific case of the peculiar AXP
4U~0142+61.   This AXP  is  by far  the best  observed  source in  the
near-infrared (NIR), optical,  and ultraviolet (UV) bands  and has two
characteristics that make it a peculiar  object. The first is that
4U~0142+61      presents     a      confirmed     infrared      excess
\citep{2000Natur.408..689H} that  might be attributed to  an accretion
disk, whereas the second one is that  it is too bright for its cooling
age, thus challenging  the conventional magnetar model.   Here we show
that the properties  of this AXP can  be well explained by  a model in
which the central compact remnant  is a massive magnetized white dwarf
resulting  from the  merger of  two otherwise  ordinary white  dwarfs,
surrounded by the heavy accretion disk produced during the merger.

\section{A model for 4U~0142+61}

To  start  with,  we  compute  the  approximate  mass  and  radius  of
4U~0142+61. The  stability of general relativistic  uniformly rotating
white  dwarfs has  been recently  studied \citep{2013ApJ...762..117B},
and  it  has  been  shown  that  constant  rotation  period  sequences
intersect  the stability  region of  white dwarfs  in two  points that
determine lower and upper bounds  for the mass, equatorial/polar radii
and moment of  inertia.  In Table~\ref{tab:bounds} we  show the bounds
for 4U~0142+61.  In this  table $\langle R\rangle=(2 R_{\rm eq}+R_{\rm
p})/3$ denotes  the mean-radius,  where $R_{\rm  eq}$ and  $R_{\rm p}$
are, respectively, the equatorial and polar radii.

\begin{deluxetable}{lcc}[t!]
\tabletypesize{\scriptsize} 
\tablecaption{Bounds  for the mass,  radius and  moment of  inertia of
  4U~0142+61.
\label{tab:bounds}}
\tablewidth{0pt}
\tablehead{
\colhead{} &
\colhead{Minimum} &
\colhead{Maximum}
 }
\startdata
$M\; (M_{\sun})$                 & 1.16                & 1.39 \\
$R_{\rm eq}\; (10^8$~cm)         & 1.05                & 6.66 \\
$\langle R \rangle \, (10^8$~cm) & 1.05                & 6.03 \\
$I$~(g~cm$^2$)                   & $2.9\times 10^{48}$ & $1.4\times 10^{50}$ 
\enddata
\end{deluxetable}

\subsection{IR, optical and UV photometry}
\label{sec:photometry}

We  next  fitted  the  spectrum  of  4U~0142+61  as  the  sum  of  two
components. The first is a black body:
\begin{equation}
F_{\rm BB} = \pi 
\frac{2 h}{c^2} \left( \frac{R_{\rm WD}}{d} \right)^2 
\frac{\nu^3}{e^{h \nu/(k_{\rm B} T_{\rm eff})}-1}\, ,
\end{equation}
where $R_{\rm WD}$ and $T_{\rm eff}$ are, respectively, the radius and
effective  temperature of  the white  dwarf. As  it is shown below  in
Sect.~\ref{sec:mag-age},  the  system  now   behaves  as  an  ejector,
inhibiting the accretion  of the disk material onto  the central white
dwarf.  Thus, for the second component  we adopted the black body disk
model  of \cite{1997ApJ...490..368C},  which is  more appropriate  for
these systems \citep{2007ApJ...661L.179G}:
\begin{eqnarray}
F_{\rm disk}=&\, & 12 \pi^{1/3} \cos i 
\left( \frac{R_{\rm WD}}{d} \right)^2 
\left( \frac{2 k_{\rm B} T_{\rm eff}}{3 h \nu} \right)^{8/3} 
\left( \frac{h \nu^3}{c^2} \right) \nonumber\\
& \, &\times \int_{x_{\rm in}}^{x_{\rm out}} \frac{x^{5/3}}{e^x-1}dx\, ,
\end{eqnarray}
where $i$ is the inclination angle of  the disk, which we assume to be
face-on,  and  $x=h  \nu/(k_{\rm  B}  T)$.  In  this  model  the  disk
temperature $T$ varies as $r^{-3/4}$ \citep{1997ApJ...490..368C}, with
$r$  the distance  from the  center of  the white  dwarf. It  is worth
mentioning     that    in     previous    studies     of    4U~0142+61
\citep{2000Natur.408..689H,  2006Natur.440..772W} the  irradiated disk
model of  \cite{1990A&A...235..162V} has  been used instead,  but this
model is more appropriate for accreting sources.

\begin{figure}[t]
\centering
\includegraphics[width=\columnwidth,clip]{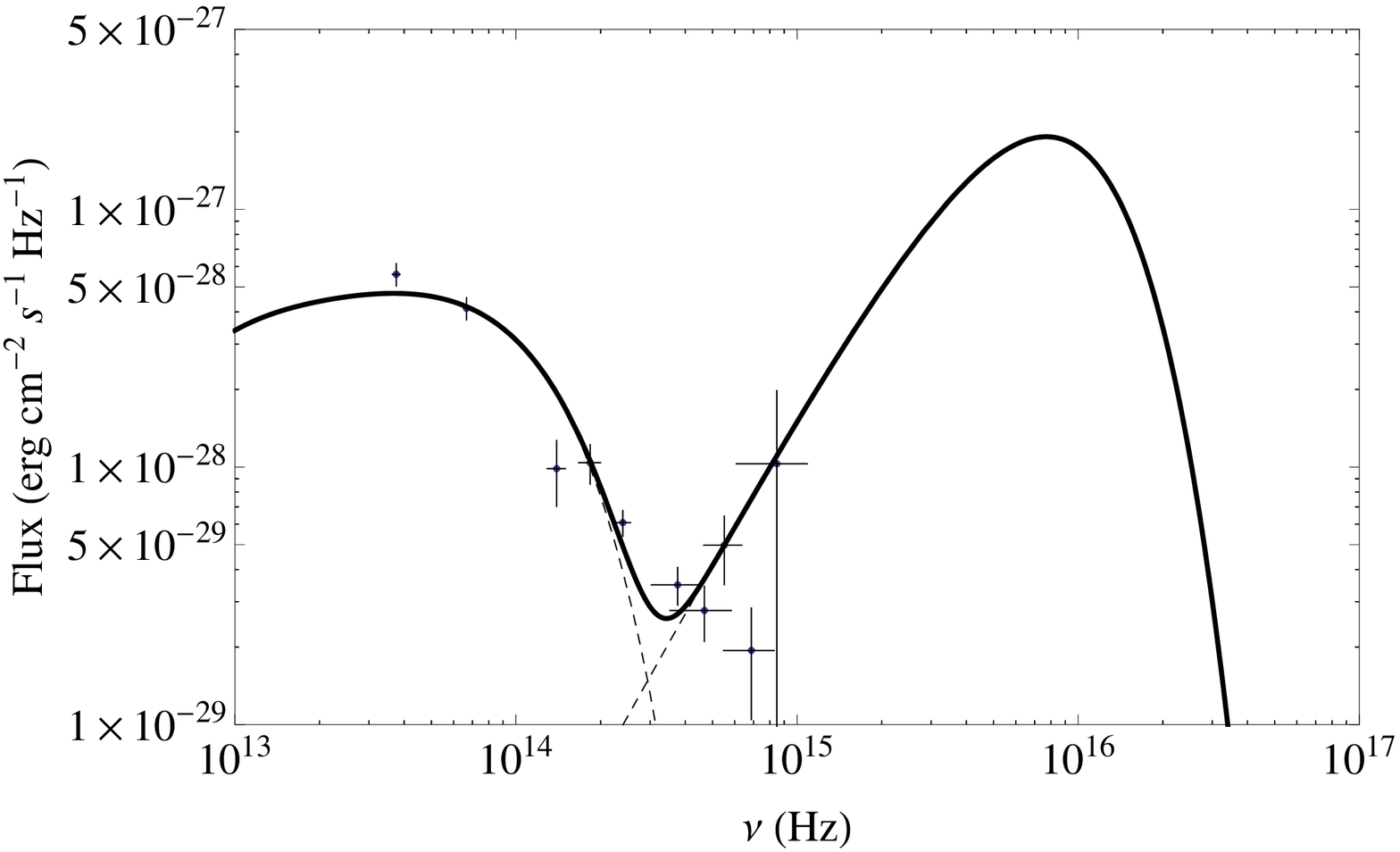}
\caption{Observed and fitted  spectrum of 4U~0142+61. Due  to the high
  variability of  the source in the  optical bands we average  all the
  existing data of the source in  the different bands.  All these data
  come  from observations  from 31  October 1994  up to  26 July  2005
  \citep{2000Natur.408..689H,                     2004A&A...416.1037H,
  2005MNRAS.363..609D,    2005AdSpR..35.1177M,    2006ApJ...652..576D,
  2009PASJ...61...51M}.  The result of the  average is $V = 25.66$, $R
  = 25.25$, $I = 23.76$, $J =  22.04$, $H = 20.70$, $K = 19.97$. There
  are  upper   limits  in  the   $U$  and   $B$  bands,  $U   =  25.8$
  \citep{2005MNRAS.363..609D}        and       $B        =       28.1$
  \citep{2004A&A...416.1037H},  respectively.   We also  consider  the
  observations  of  \cite{2006Natur.440..772W} with  the  Spitzer/IRAC
  instrument at wavelengths 4.5~$\mu$m  and 8.0~$\mu$m. The fluxes are
  36.3~$\mu$Jy and 51.9~$\mu$Jy, respectively.   We corrected the data
  for  the  interstellar  extinction,  using  the  estimated  distance
  $d=3.6$~kpc \citep{2006ApJ...650.1070D} and an absorption in the $V$
  band  $A_V=3.5$ \citep{2006ApJ...650.1082D}.   For the  rest of  the
  bands  we used  $A_U=1.569 A_V$,  $A_B=1.337 A_V$,  $A_R=0.751 A_V$,
  $A_I=0.479 A_V$, $A_J=1.569 A_V$,  $A_J=0.282 A_V$, $A_H=0.190 A_V$,
  and $A_K=0.114 A_V$  \citep{1989ApJ...345..245C}.  The extinction in
  the  Spitzer/IRAC bands  for $A_K<0.5$  are $A_{4.5\mu{\rm  m}}=0.26
  A_K$ and $A_{8.0\mu{\rm m}}=0.21 A_K$ \citep{2009ApJ...690..496C}.}
\label{fig:fit4U}
\end{figure}

We computed the  best-fit of the spectrum  parameters, finding $R_{\rm
WD}\approx 0.006\, R_{\sun}$,  $T_{\rm eff}\approx 1.31\times 10^5$~K,
inner  and outer  disk radii  $R_{\rm in}=0.97  \, R_{\sun}$,  $R_{\rm
out}=51.1\,  R_{\sun}$  and  correspondingly   inner  and  outer  disk
temperatures  $T_{\rm  in}\approx   1950$~K  and  $T_{\rm  out}\approx
100$~K, respectively. In Fig.~\ref{fig:fit4U}  we show the photometric
data of 4U~0142+61 and our best-fit composite spectrum.  The agreement
of the composite  spectrum with the observational data  is quite good,
taking into account  that the high variability of the  source in these
bands can lead to changes in the  optical fluxes of up to one order of
magnitude \citep{2006ApJ...652..576D}. Thus, the  white dwarf model is
compatible with the observed photometry  of 4U~0142+61, as it seems to
occur   for   SGR~0418+5729,  Swift~J1822.3--1606,   and   1E~2259+586
\citep{BIRR2013}.

To  check  whether  this  is  a realistic  and  consistent  model  for
4U~0142+61, we  ran a SPH  simulation of the  merger of a  $0.6+1.0 \,
M_{\sun}$ binary  white dwarf, which  results in a central  remnant of
$\approx  1.1\, M_{\sun}$,  with a  radius $R_{\rm  WD}\approx 0.006\,
R_{\sun}$, in  agreement with the  photometric value.  We  recall that
the  central  white   dwarf  accretes  some   material  from  the
surrounding disk  (of mass $M_{\rm disk}\approx  0.5\, M_{\sun}$) and,
thus,  shrinks  a  little.   Moreover,  the  rotation  period  is
$P\approx 15.7$~s and the moment of inertia of the central white dwarf
and the hot corona is  $I\approx 2.0\times 10^{50}$~g~cm$^2$, which is
slightly    larger    than    our    maximum    estimate    ---    see
Table~\ref{tab:bounds}.  Furthermore, the  magnetic field generated in
the differentially-rotating  hot corona  produced in the  aftermath of
the  merger amounts  to  $B\approx  10^{10}$~G \citep{enrique2012}  which
amply    explains   the    magnetic   field    of   4U~0142+61,    see
Sect.~\ref{sec:mag-age}.

\subsection{The age and magnetic field of 4U~0142+61}
\label{sec:mag-age}

\begin{figure}[t]
\centering
\includegraphics[width=0.8\columnwidth,clip]{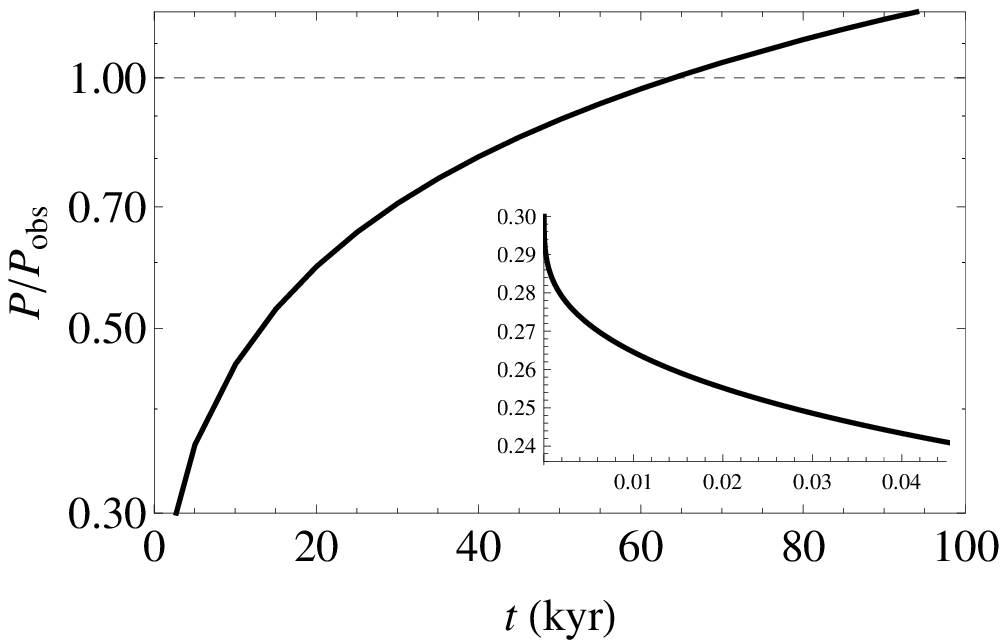}\\
\includegraphics[width=0.8\columnwidth,clip]{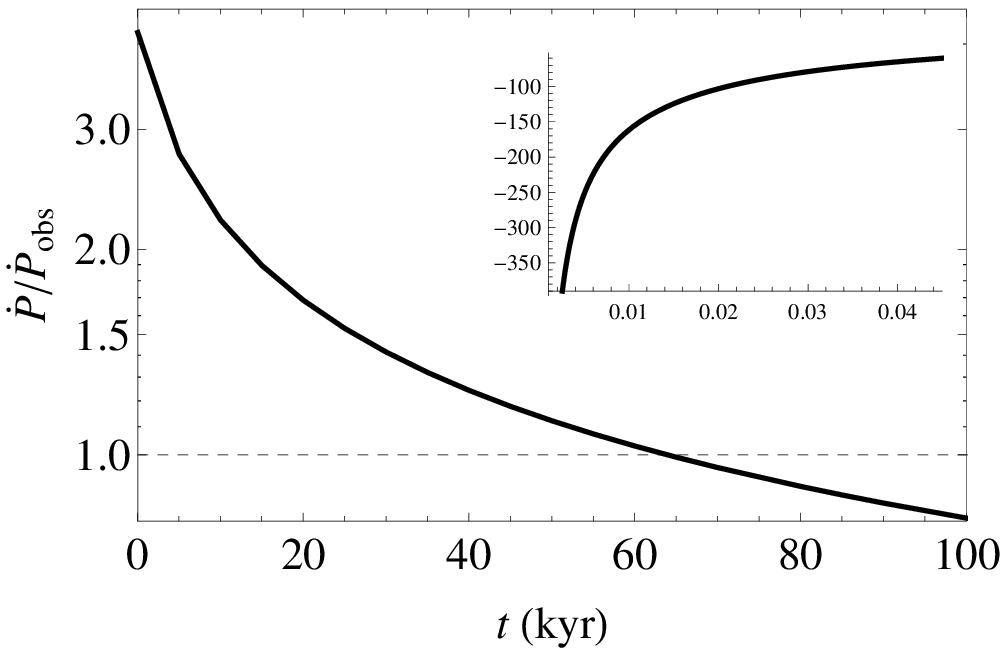}
\caption{Time  evolution   of  the  period  (top   panel)  and  period
  derivative  (bottom panel)  of 4U~0142+61.  The insets  show  to the
  early evolutionary phases of the system.}
\label{fig:ppdot}
\end{figure}

The presence of  a disk around the magnetized white  dwarf plays a key
role in  the evolution of  its rotation  period.  This results  from a
delicate  interplay  between the  interaction  of  the disk  with  the
magnetosphere  of the  star, and  accretion  of disk  matter onto  the
surface of  the white dwarf.   A solution of  the magneto-hydrodynamic
equations including the explicit coupling of magnetosphere-disk system
and the mass  and angular momentum transfer from the  disk to the star
is not yet  available.  For this reason the torque  acting on the star
it is often followed using a phenomenological treatment.  We adopt the
model of  \cite{1996MNRAS.280..458A}, which assumes that  the magnetic
field  lines  threading  the  disk  are closed.   In  this  model  the
evolution of $\omega$ is dictated by
\begin{eqnarray}
\label{eq:spindown}
\dot{\omega}=&-&\frac{2 B^2 \langle R\rangle^6 \omega^3}{3 I c^3} 
\sin^2\theta + 
\frac{B^2 \langle R\rangle^6}{3 I}\left[ \frac{1}{R^3_{\rm mag}}-
\frac{2}{(R_{\rm c} R_{\rm mag})^{3/2}} \right]\nonumber\\
&+&\frac{\dot{M} R^2_{\rm mag} \omega}{I}\, ,
\end{eqnarray}
where $\theta$ is the angle between the rotation axis and the magnetic
dipole  moment,   $R_{\rm  mag}   =[B^2  \langle{R}\rangle^6/({\dot{M}
\sqrt{2    G    M}})]^{2/7}$     is    the    magnetospheric    radius
\citep{2000ApJ...534..373C,2012MNRAS.420..810T,   2012ApJ...745..101M,
2012ApJ...758L...7R},  and $R_{\rm  c}= (G  M/\omega^2)^{1/3}$ is  the
corotation radius. The first term in Eq.~(\ref{eq:spindown}) describes
the  traditional  magneto-dipole  braking,   the  second  one  is  the
star-disk coupling, while the last  one describes the angular momentum
transfer from the disk to the white dwarf.  We adopt an accretion rate
corresponding  to a  Shakura-Sunyaev viscosity  parameter $\alpha_{\rm
SS}=0.1$        \citep{1990ApJ...351...38C,       2000ApJ...534..373C,
2009ApJ...702.1309E}.   Adopting a  misalignment angle  $\theta=\pi/2$
and integrating Eq.~(\ref{eq:spindown}) using the parameters resulting
from  our SPH  simulation and  it results  that, for  a wide  range of
magnetic field  strengths, at early stages  $R_{\rm mag}\approx R_{\rm
WD}$.  Thus, initially the star is  spun-up due to the large accretion
rates  --- see  the  insets of  Fig.~\ref{fig:ppdot}.  However,  after
approximately $1$~kyr, the inner radius of  the disk --- which is approximately
given by  the magnetospheric  radius --- becomes  larger than  the the
light cylinder radius, $R_{\rm  lc}=c/\omega$.  Hence, the disk cannot
torque  any longer  the white  dwarf,  the rotation  period reaches  a
minimum,  and  from  this  point  on the  disk  and  the  star  evolve
independently, and accretion onto the magnetic poles stops.  Thus, the
star  behaves  as a  normal  pulsar,  spinning-down by  magneto-dipole
radiation   \citep{1973ApJ...184..271L,   2000ApJ...534..373C}.    The
surface magnetic  field needed to fit  the observed values of  $P$ and
$\dot  P$ when  a  mass  $M=1.2\, M_{\sun}$  is  adopted is  $B\approx
2.3\times   10^8$~G  at   an  age   $\tau_{\rm  sd}=64$~kyr   ---  see
Fig.~\ref{fig:ppdot}.   This  age  estimate  compares  well  with  the
spin-down characteristic  age $P/(2\dot P)\approx  68$~kyr.  Moreover,
the strength of the magnetic field  can be compared with that directly
derived using the traditional misaligned dipole expression
\begin{equation}
B=\sqrt{\frac{3 c^3 I}{\langle R\rangle^6} 
\frac{P \dot{P}}{8 \pi^2}}\, , 
\end{equation}
\noindent \citep{ferrari69, 1973ApJ...184..271L} --- which in our case
is    valid    because    $R_{\rm    in}\approx\,    R_{\sun}$    (see
Sect.~\ref{sec:photometry})  is larger  than the  radius of  the light
cylinder $R_{\rm lc}\approx 0.6\, R_{\sun}$.  From the observed values
$P=8.69$~s     and     $\dot{P}     =    2.03     \times     10^{-12}$
\citep{2000Natur.408..689H},  we  obtain  $B=2.3\times  10^{8}$~G  for
$M_{\rm  min}$,  and  $6.2\times   10^{9}$~G  for  $M_{\rm  max}$,  in
agreement      with      the     result      obtained      integrating
Eq.~(\ref{eq:spindown}).   Additionally, there  are other  indications
that the magnetic field derived in  this way is sound.  In particular,
the spectrum of 4U~0142+61 exhibits a significant drop-off between the
$B$  and $V$  bands,  at  a frequency  $\nu  \approx 10^{15}$~Hz,  see
Fig.~\ref{fig:fit4U}.  \cite{2004A&A...416.1037H}  concluded that this
feature is not due to variability and, moreover, they advanced that it
is consistent with the electron cyclotron emission of a magnetic field
$B\approx 10^8$~G.  Adopting the minimum  and maximum masses derived from
our model we obtain electron  cyclotron frequencies $\nu_{\rm cyc} = e
B/(2   \pi  m_{\rm   e}   c)=6.3\times   10^{14}$~Hz  and   $1.7\times
10^{16}$~Hz, which correspond to wavelengths that fall between the NIR
and the UV, 0.5 and  176~$\mu$m, respectively.  This suggests that the
magnetic field must  be closer to the lower value,  and therefore that
the corresponding mass should be approximately $1.2~M_{\sun}$.  Actually, it
is interesting to realize that although the mass of the remnant of the
coalescence is slightly smaller than  our fiducial mass for 4U~0142+61
--- $\approx 1.1$  and $\approx 1.2\,  M_{\sun}$, respectively ---  the mass
accreted  during  the  spin-up  phase  is  $M_{\rm  acc}  \approx  0.05\,
M_{\sun}$,  in  good   agreement  with  the  mass   derived  from  the
photometric solution.   It could  be argued that  this is  the maximum
possible   accreted   mass,   since    during   these   early   stages
super-Eddington accretion rates are needed to accrete all the material
inflowing from the  disk.  Nevertheless, during the  very early stages
after the merger the temperature of the coalesced system is very high,
and the emission of neutrinos is not negligible \citep{enrique2012}.

We now compute the cooling age  of 4U~0142+61, and compare it with the
spin-down  age.  As  the  hot, convective  corona  resulting from  the
merger is  very short-lived  \citep{enrique2012} the evolution  of the
surface luminosity of the white  dwarf can be estimated using Mestel's
cooling law \citep{1952MNRAS.112..583M}:
\begin{equation}
\label{eq:tau_cool}
\tau_{\rm cool}=\frac{1}{\langle A\rangle}
\left(\frac{b M Z^{2/5}}{L_{\rm WD}} \right)^{1/x}
-0.1\, ,
\end{equation}
where $\langle A\rangle$  is the average atomic weight of  the core of
the  white dwarf,  $Z$ is  the metallicity  of its  envelope, $x=1.4$,
$b=635$ \citep{2003ApJ...589..179H}, and the  rest of the symbols have
their  usual  meaning. Adopting  a  carbon-oxygen  core and  $Z\approx
0.001$, which  is a  reasonable value  \citep{2010A&ARv..18..471A}, we
obtain  a  cooling  age   $\tau_{\rm  cool}\approx  64$~kyr,  in  good
agreement with the spin-down age.

\subsection{X-ray luminosity}
\label{sec:x-ray}

For  a distance  $d=3.6$~kpc, \cite{2006ApJ...650.1070D}  estimated an
isotropic  X-ray  luminosity  $L_X=4\pi d^2  f^{\rm  unabs}_X  \approx
1.3\times  10^{35}$~erg~s$^{-1}$,  using  the  unabsorbed  X-ray  flux
$f^{\rm  unabs}_X=8.3\times 10^{-11}$~erg~s$^{-1}$~cm$^{-2}$  obtained
by  \cite{2003ApJ...587..367P}.  We  use  the  result  of  the  latest
observations  of  4U~0142+61  with   the  EPIC  cameras  onboard  {\sl
XMM-Newton},                 $f^{\rm                unabs}_X=7.2\times
10^{-11}$~erg~s$^{-1}$~cm$^{-2}$          \citep{2005A&A...433.1079G},
obtaining  $L_X\approx 1.1\times  10^{35}$~erg~s$^{-1}$ when  the same
distance is  adopted.  The loss  of rotational energy  associated with
the  spin-down  of  4U~0142+61,  $\dot{E}_{\rm   rot}  =  -4  \pi^2  I
\dot{P}/P^3$        gives        $|\dot{E}_{\rm        rot}|=1.7\times
10^{37}$~erg~s$^{-1}$    for    $M_{\rm     min}$    and    $3.4\times
10^{35}$~erg~s$^{-1}$  for $M_{\rm  max}$,  that  cover the  estimated
X-ray luminosity.

The time integrated X-ray spectrum  of 4U~0142+61 is well described by
a  black body  and a  power-law  model with  $k_{\rm B}  T_{\rm BB}  =
0.4$~keV and photon index $\Gamma = 3.62$ \citep{2005A&A...433.1079G}.
The black body component corresponds  to a temperature $T_{\rm BB}\approx
4.6\times 10^6$ K,  which is higher than the surface  temperature of a
hot white dwarf. However, these  systems may have coronal temperatures
much higher than that  of the surface \citep{2012PASJ...64...56M}, and
thus the  X-ray emission would  be of magnetospheric  origin.  Because
the inner radius  of the disk is  larger than the radius  of the light
cylinder $R_{\rm  lc}$ (see  Sect.  \ref{sec:mag-age})  the mechanisms
producing  such  radiation  are  similar  to  those  of  pulsars.   In
particular, a possible mechanism  was delineated by \cite{usov93}, who
showed  that reheating  of  the magnetosphere  by  the bombardment  of
positrons moving backward to the surface of the star can produce large
X-ray  luminosities.   Positrons  are   produced  the  interaction  of
high-energy  photons with  ultra-relativistic electrons,  resulting in
the  creation  of  electron-positron  pairs.   Following  closely  the
calculations of  \cite{usov93} we computed the  theoretically expected
X-ray luminosity of 4U~0142+61.  We  found that the reheating of polar
caps  produces   a  persistent   X-ray  luminosity   $L_X\approx  2\times
10^{35}$~erg~s$^{-1}$, in  agreement with  observations.  Nonetheless,
there  are   other  possibilities.    If  the   conventional  magnetar
interpretation is  adopted, the X-ray  luminosity would be due  to the
neutron  star.   Alternatively,  it  could  also  be  due  to  ongoing
accretion from a fossil disk  onto the neutron star \citep{Alpar2001}.
In  such cases  the  white  dwarf product  of  the  merger would  have
accreted enough  material to undergo  accretion induced collapse  to a
neutron star.

\section{Conclusions}

We  studied the  possibility that  the  peculiar AXP  4U~0141+61 is  a
massive, fast-rotating, highly magnetized white dwarf, and we explored
the viability of this object being  the result of the coalescence of a
binary  white  dwarf.   Specifically,  from  its  observed  rotational
velocity we first  derived bounds for the mass, radius,  and moment of
inertia.   Afterwards, we  fitted  the  IR, optical,  and  UV data  of
4U~0142+61 with a  composite spectrum made of two  components, a black
body and a dust disk, finding  a good agreement with the observations.
Moreover,  we  showed  that  the   characteristics  of  the  disk  are
consistent with  the results of  a SPH simulation  of the merger  of a
$0.6 + 1.0\,  M_{\sun}$ binary system.  We then estimated  the age and
the  magnetic field  of this  AXP.  Adopting  the results  of our  SPH
simulation we  obtained a magnetic  field $B=2.3\times 10^8$~G,  and a
post-merger  age $\approx  64$~kyr.  The  cyclotron frequency  of this
magnetic field $\nu_{\rm cyc}\approx 6\times 10^{14}$~Hz would explain an
absorption feature observed in the  spectrum of 4U~0142+61 at $\nu\approx
10^{15}$~Hz.  Furthermore, our age  estimate is in excellent agreement
with  the white  dwarf cooling  age.  We  also showed  that the  X-ray
luminosity  of 4U~0142+61  can  be well  explained  by the  rotational
energy  loss,  and  we  inferred a  theoretical  estimate  $L_X\approx
2\times 10^{35}$~erg~s$^{-1}$,  which agrees with the  observed value,
$L_X\approx1.09 \times 10^{35}$~erg~s$^{-1}$.   All these findings may
support the hypothesis that the peculiar AXP 4U~0141+61 was originated
in a white dwarf binary merger.

\acknowledgements  

This work  was partially supported  by MCINN grant  AYA2011--23102, by
the  AGAUR,  by  the  European  Union FEDER  funds,  and  by  the  ESF
EUROGENESIS project (grant EUI2009-04167).

\end{document}